\documentclass[sigconf]{templates/acmart}

\usepackage{hyperref}

\usepackage[table]{xcolor}
\usepackage[utf8]{inputenc}
\usepackage{tabularx}   
\usepackage{booktabs}   
\usepackage{caption}    
\usepackage{ragged2e}   
\usepackage{adjustbox}
\usepackage{tcolorbox}

\usepackage{titlesec}
\AtBeginDocument{%
  }

\titlespacing*{\subsection}{0pt}{0.6em}{0.3em}
\titlespacing*{\section}{0pt}{0.6em}{0.6em}

\setcopyright{none}
\copyrightyear{}
\acmISBN{}
\renewcommand\footnotetextcopyrightpermission[1]{}
\acmConference{}{}

\usepackage{enumitem}
\usepackage{tcolorbox}
\usepackage{hyperref}
\usepackage[dvipsnames]{xcolor}
\usepackage{graphicx}

\usepackage[T1]{fontenc}
\usepackage{breakurl} 

\begin{document}

\title{Doc To The Future: Infomorphs for Interactive, Multimodal Document Transformation and Generation}

\author{Balasaravanan Thoravi Kumaravel}
\email{bala.kumaravel@microsoft.com}
\affiliation{%
  \institution{Microsoft Research}
  \city{Redmond}
  \state{Washington}
  \country{USA}
}

\newcommand{\systemname}{\texttt{DocuCraft~}}
\newcommand{\systemnamedot}{\texttt{DocuCraft}}
\newcommand{\bala}[1]{\textcolor{red}{#1}}
\definecolor{darkgreen}{rgb}{0.13, 0.55, 0.13}
\definecolor{darkgrey}{rgb}{0.25, 0.25, 0.25}

\newcommand{\filename}[1]{%
  \textcolor{darkgreen}{%
    \begingroup
    \ttfamily
    \hyphenchar\font=`\-
    #1%
    \endgroup
  }%
}

\newcommand{\imnode}[1]{\texttt{\textcolor{darkgrey}{#1}}}
\newcommand{\linktext}[1]{\texttt{\textcolor{blue}{#1}}}

\newcommand{\scenarioheading}[1]{%
  {\fontfamily{bch}\selectfont
  \vspace{5pt}
  \noindent\textbf{#1}}%
}
\begin{abstract}
Creating new documents by synthesizing information from existing sources is an important part of knowledge work in many domains. This process often involves gathering content from multiple documents, organizing it, and then transforming it into new forms such as reports, slides, or spreadsheets. While recent advances in Generative AI have shown potential in automating parts of this process, they often provide limited user control over the handling of multimodal inputs and outputs. In this work, we introduce the notion of ``\textit{infomorphs}'' which are modular, user-steerable, AI-augmented transformations that support controlled synthesis, and restructuring of information across formats and modalities. We propose a design space that leverage \textit{infomorph}-driven workflows to enable flexible, interactive, and multimodal document creation by combining Generative AI techniques with user intent and desired information context. As a concrete instantiation of this design space, we present \systemnamedot, a canvas-based interface to visually compose \textit{infomorph} workflows.  \systemname allows users to chain together \textit{infomorphs} that perform operations such as page extraction, content summarization, reformatting, and generation, leveraging Generative AI at each stage to support rich, cross-document and cross-modal transformations. We demonstrate the capabilities of \systemname through an example-driven usage scenario that spans across different facets of common knowledge work tasks illustrating its support for fluid, human-in-the-loop document synthesis and highlights opportunities for more transparent and modular interaction for Generative AI-assisted information work.

\end{abstract}


\begin{teaserfigure}
  \fbox{\includegraphics[width=\textwidth]{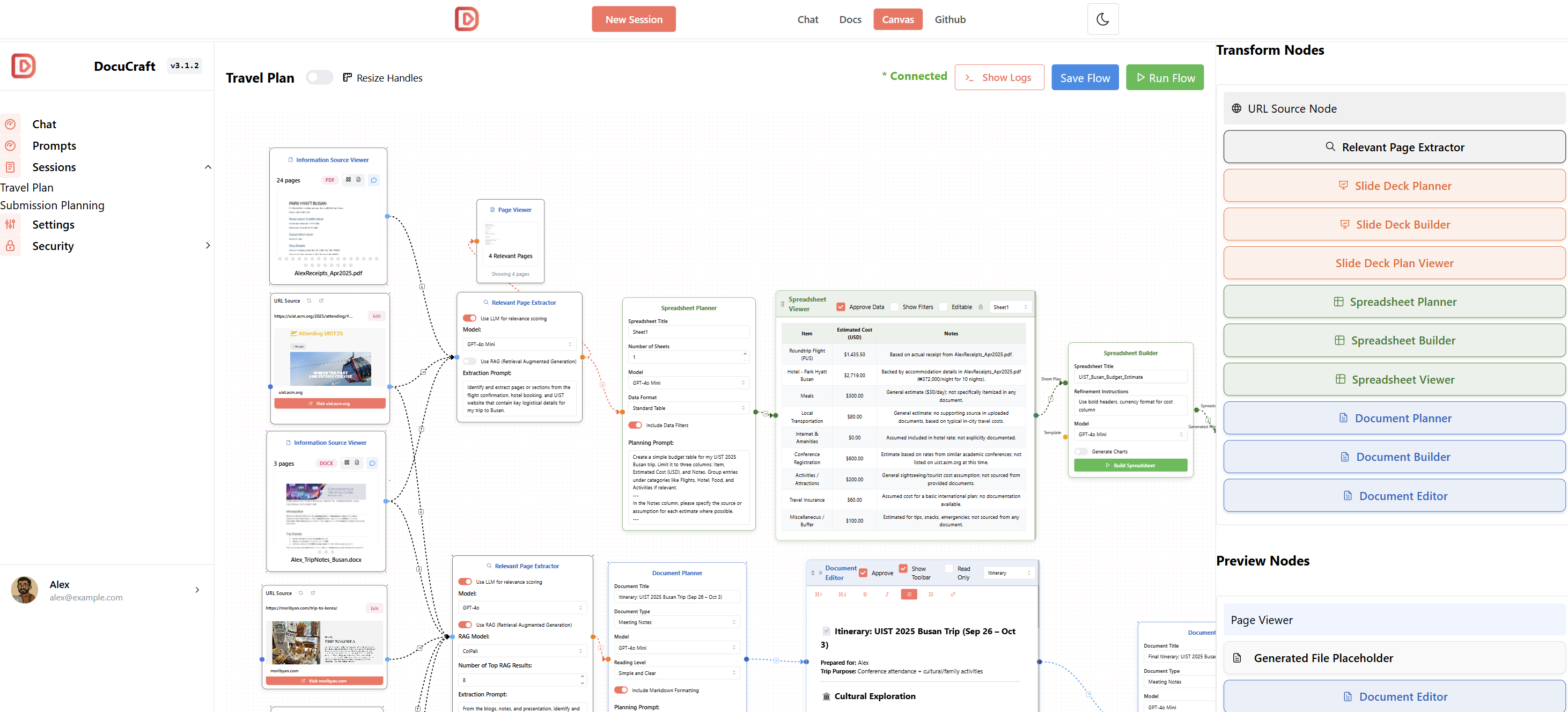}}
  \caption{The \systemname interface: Users visually compose document transformation workflows by connecting nodes representing sources, AI-driven operations (Infomorphs like Planners and Extractors), and viewers for intermediate results, enabling controlled synthesis. Supplemental demo video  can be found \href{https://www.tkbala.com/docucraft_yt}{\linktext{here}}.}
  \Description{}
  \label{fig:teaser}
\end{teaserfigure}

\settopmatter{printfolios=true}

\maketitle

\section{Introduction}

In today's professional world, knowledge work plays a critical role in innovation and productivity across many disciplines. During such work, information workers search, gather, and organize information from a wide variety of sources, whether for preparing structured reports, designing presentations, or writing articles. This entire process is rarely linear and often involves iterations of exploration of information sources, rating them as relevant and evaluating them as significant. These behaviors reflect a strategic approach that is both goal driven and resource aware. 

Historically, the primary challenge often lay in efficiently retrieving relevant documents. Search tools and engines significantly reduced this retrieval effort, allowing users to access vast amounts of information with minimal input \cite{10.1145/3209978.3210027}. However, as access to relevant information became easier, the focus shifted towards making sense of the content and synthesizing it effectively. With the emergence of Transformer models and large language models (LLMs) capable of understanding and generating natural language, user effort has been further reduced not only for retrieving information but also for extracting, summarizing, and synthesizing it \cite{cambon2023early, slobodkin2023summhelper, serajeh2024llms, pang2025understanding}. Modern search engines have started to take advantage of these capabilities, offering a richer experience. This evolution is evident in platforms such as Perplexity\footnote{https://www.perplexity.ai/}, Google Search\footnote{https://blog.google/products/search/generative-ai-search/}, and Bing Copilot\footnote{https://copilot.microsoft.com/}, which go beyond linking to relevant content by supporting follow-up queries, contextual exploration, and more nuanced reasoning over user intent. This has enabled a deeper, conversational approach to information seeking and synthesis and has spurred renewed interest in studying how people work with information sources \cite{srinivasa2024revisiting, yun2025generative}.

Despite these advances, many current generative AI systems supporting knowledge work remain largely \textit{monolithic} and \textit{opaque}. They often operate as black-boxes, producing entire outputs in a single step, with limited visibility or control over the intermediate stages. This lack of transparency and steerability limits the user’s ability to selectively guide, inspect, or iteratively refine the information transformation process, especially in complex knowledge work where intentional, step-by-step control is critical. While template-based document generation techniques\footnote{https://automateddocs.com} have offered structured solutions for efficiency and consistency, particularly in technical or business documentation, their inherent rigidity limits flexibility when working with unstructured or multimodal content. The challenge, therefore, lies in harnessing the power of AI automation without sacrificing the user agency necessary for nuanced, complex synthesis tasks.

To address these limitations and bring clarity to the design of interactive AI systems for knowledge work, this paper proposes a design space that provides a useful lens to characterize, reason, and compare how generative AI systems can effectively support human-in-the-loop knowledge work. The design space has five core dimensions - (1) \textit{Perceptual Cues and Information Scent}, (2) \textit{User Agency and Workflow Guidance}; (3) \textit{Interaction Granularity and Structural Control}; (4) \textit{Infomorphs}, and (5) \textit{Input and Output Modalities}. This framework is intended to help move tools beyond \textit{monolithic}, \textit{black-box} generation toward more flexible, transparent, and human-in-the-loop approaches to document synthesis and transformation.

Central to enabling this flexibility, we introduce the concept of \textit{infomorphs}: modular, AI-augmented operations that allow users to compose and guide document synthesis across diverse formats and modalities. As a specific instantiation of this design space and these ideas, we showcase \systemnamedot, a system that uses a Canvas-based interface to support the visual composition of \textit{Infomorphs}. \systemname allows users to define and manipulate flows that leverage the Generative AI techniques of today on a wide variety of document types. Using the proposed design space, we compare and contrast \systemnamedot's capabilities with existing state-of-the-art tools. In our work, we specifically focus on the generation of new documents from a collection of heterogeneous, multimodal input sources, such as Word Documents, PDFs, Web pages, spreadsheets, or slide decks. Alongside single-shot generation and iterative prompting, we propose an approach of interactive, step-wise flows for transforming information, where infomorphs serve as the core conceptual unit, allowing users to guide how content is extracted, summarized, reorganized, and composed into new formats. These transformations, or infomorphs, serve as the core conceptual unit of the system through which we show how Generative AI can augment information work while preserving user agency. 

We use \systemname as a case study to demonstrate different ways, in which users can work with Generative AI while also staying in control of their workflows. Specifically, we illustrate how users can visually construct transformation pipelines by chaining together infomorphs, each representing a distinct operation such as extraction, summarization, classification, or synthesis. These modular units not only provide transparency into intermediate results, but also enable selective editing and recomposition, supporting a more iterative and exploratory process. We show how the system supports a fluid interplay between user intention and machine capability. Through this exploration, we highlight the potential of such infomorph-based interfaces to bridge the gap between end-user goals and the underlying power of large language models, fostering new patterns of interactive document construction that are explainable, adaptable, and deeply collaborative. In summary, this paper makes the following contributions.

\begin{itemize}
    \item A design space for generative AI systems that support human-in-the-loop document synthesis. This space is defined by five core dimensions:  (1) Perceptual Cues and Information Scent, (2) User control and workflow involvement, (3) Interaction granularity and structural agency, (4) input and output modalities, (5) Types of Information Transformations aka ``\textit{Infomorphs}''
    \item \systemnamedot, a canvas-based system that enables users to visually compose and manipulate infomorphs across heterogeneous, multimodal input sources.
\end{itemize}
\section{Related work}

Understanding how to design effective tools for knowledge work requires grounding in established theories of information seeking, reuse, and interaction, as well as an understanding of the evolution of supporting technologies and user strategies.

\subsection{Information Foraging and Reuse}

The strategic, goal-driven, and resource-aware nature of knowledge work aligns closely with prior research on information behavior and knowledge reuse. A foundational perspective is offered by Information Foraging Theory (IFT)~\cite{ift95}, which models information seeking as analogous to animal foraging. Users aim to maximize the value of information gained while minimizing the cost (time, cognitive effort) of acquiring it. Central to IFT is the concept of ``information scent''~\cite{informationscent, 10.1145/2600428.2609626}, which represents the user's subjective assessment of the value versus cost associated with pursuing an information source or path, based on perceptual cues like summaries, titles, or previews. These cues guide users' decisions about whether to delve deeper into a source or continue searching elsewhere. Complementing IFT, research on knowledge reuse examines how people adapt and re-purpose existing content to efficiently meet new goals. For instance, Domain Theory \cite{sutcliffe2002domain} emphasizes the value of structuring prior knowledge into reusable models, a principle relevant in content creation and software development \cite{10.1145/3449240}. Together, these theories underscore the importance of designing systems that provide effective cues (scent) and support the efficient reuse and transformation of information.

\subsection{User Strategies and Interaction Cues}

Building on these theoretical foundations, researchers have extensively studied how users interact with information environments. Studies have shown how users leverage information cues, such as headings, summaries, previews, and link text, to quickly assess the relevance of information sources before committing to more effortful engagement \cite{whatareyoulookingfor, 10.1145/365024.365331, 10.1145/3341981.3344231}. Information scent guides navigation patterns, particularly on the Web, where users follow links based on their perceived relevance to their evolving information needs \cite{informationscent, 10.1145/2600428.2609626}. As information retrieval became less effortful due to improved search technologies \cite{10.1145/3209978.3210027}, user strategies adapted. The bottleneck shifted from finding sources to making sense of the content within them. Users developed sophisticated strategies to optimize how they switch between skimming and deep reading, and how they synthesize information across multiple documents, always driven by the underlying goal of maximizing knowledge throughput while minimizing effort. This highlights the need for tools that support not just finding information, but also processing, evaluating, and integrating it effectively.

\subsection{Evolution of Tools: From Search to Generative AI}

The technological landscape supporting knowledge work has evolved significantly. Early search engines dramatically lowered the barrier to accessing relevant documents \cite{10.1145/3209978.3210027}. More recently, the advent of powerful LLMs has further reduced the effort required for tasks beyond retrieval, such as extraction, summarization, and synthesis \cite{cambon2023early, slobodkin2023summhelper, serajeh2024llms, pang2025understanding}. Modern platforms like Perplexity, Google Search's AI Overviews, and Bing Copilot integrate these capabilities, aiming to provide direct answers, summaries, and conversational follow-ups. These systems attempt to enhance the perceived "information scent" by offering synthesized information upfront, potentially enabling more seamless navigation of complex information spaces \cite{srinivasa2024revisiting, yun2025generative}. This evolution towards AI-driven synthesis has revitalized research into how people interact with information systems during knowledge work. However, the nature of "information scent" itself evolves in this context; assessing the trustworthiness and quality of AI-generated summaries or syntheses presents different challenges than assessing the relevance of a document link based on its snippet.

\subsection{Limitations of Existing Systems and Approaches}

Despite the potential of generative AI, many current systems exhibit significant limitations for complex knowledge work. A primary concern is their often \textit{monolithic} and \textit{opaque} nature \cite{yun2025generative}. They frequently function as black-boxes, taking inputs and producing complex outputs (e.g., full reports, summaries of multiple documents) in a single step. This lack of transparency into the intermediate processing stages hinders users' ability to understand how an output was derived, verify its accuracy, or selectively guide, inspect, and iteratively refine the process. This is particularly problematic for tasks requiring careful integration of information, nuanced argumentation, or adherence to specific constraints, where step-by-step control is crucial. While alternative approaches like template-based document generation offer structure and consistency, they typically lack the flexibility needed to handle diverse, unstructured inputs or to support the exploratory and iterative nature of many knowledge work tasks. The landscape of interactive AI systems remains complex \cite{yun2025generative}, with a clear need for approaches that better balance automation with user control and transparency. The opacity of current monolithic systems directly motivates exploration into modular and `inspectable' architectures.

This paper addresses the identified gaps by proposing a structured approach to designing and understanding interactive, AI-augmented systems for knowledge work. Building on foundational concepts like Information Foraging Theory and the observed limitations of current tools, our work explores how generative AI powered workflows can enhance information processing beyond initial retrieval, focusing on the transformation and synthesis stages. While existing approaches offer valuable insights, the specific challenges of balancing generative AI's power with user control in flexible document creation workflows necessitate a more structured framework. To systematically explore how systems can best support this process, we propose a design space grounded in practical exploration and the core operational concept of modular transformations - \textit{infomorphs}.
\section{Design Space of Generative Document Workflows}

\begin{table*}[t]
\centering
\captionsetup{justification=centering}
\caption{Comparison of DocuCraft and related AI systems across the five proposed design dimensions for document-centric information work}
\renewcommand{\arraystretch}{1.5}
\rowcolors{2}{gray!10}{white}
\begin{adjustbox}{max width=\textwidth}
\begin{tabular}{>{\raggedright\arraybackslash}p{1.7cm} 
                >{\raggedright\arraybackslash}p{2.8cm} 
                >{\raggedright\arraybackslash}p{3.2cm} 
                >{\raggedright\arraybackslash}p{3.2cm} 
                >{\raggedright\arraybackslash}p{3.2cm} 
                >{\raggedright\arraybackslash}p{2.8cm}}
\toprule
\textbf{Design Dimension} & 
\textbf{\systemname} \newline (Our System) & 
\textbf{Standard LLM Chatbots} \newline (e.g., ChatGPT, Gemini) & 
\textbf{AI-Enhanced Search} \newline (e.g., Perplexity, Google) & 
\textbf{Integrated Assistants} \newline (e.g., M365 Copilot) & 
\textbf{Template-Based Automation} \\
\midrule
D1 & Rich previews, summaries, source chat (High support) 
   & Limited (source links often absent/basic) 
   & Source links, snippets, summaries (Medium) 
   & Varies by app; source context often implicit 
   & N/A (Uses pre-defined inputs) \\
   
D2 & High (User composes visual workflow; allows step inspection) 
   & Medium (Prompt iteration, chain-of-thought) 
   & Low (System generates overview/answer) 
   & Medium (Commands within app context) 
   & Low (Follows rigid template) \\
   
D3 & High (Intent, Format, Co-Structure, Full Edit) 
   & Medium (Intent, Format via prompt) 
   & Low (System defines structure) 
   & Medium (Operates on app structures) 
   & High (User defines template precisely) \\
   
D4 & Explicit \& Composable (Scatter, Gather, Transduce) 
   & Implicit/Monolithic (Primarily Gather/Transduce via prompt) 
   & Monolithic (Gather/Summarize focus) 
   & Task-specific commands (Low Modularity) 
   & Pre-defined steps (Low AI Modularity) \\
   
D5 & Rich Multimodal (PDF, DOCX, PPTX, XLS, Web, Image $\rightarrow$ PPTX, XLSX, DOCX) 
   & Primarily Text (Emerging Image/Data Input/Output) 
   & Primarily Web/Text $\rightarrow$ Text Summary 
   & App-dependent (e.g., Word=Text, PPT=Slides) 
   & Often Text/Structured Data \\
\bottomrule
\end{tabular}
\end{adjustbox}
\end{table*}

The creation of new documents from existing information sources is a key aspect of knowledge work, and modern transformer-based language models offer significant potential in this area due to their text and visual understanding capabilities. To structure our exploration of how these capabilities can best support users, we introduce a core operational concept:

\textsf{\textit{infomorphs}: modular, AI-augmented transformations that users can flexibly compose to guide document synthesis across diverse formats and modalities. Infomorphs provide discrete, composable units of information transformation, balancing user intent with AI-driven automation.}

The need for such a concept, and the broader design space, emerged from an iterative development process. Early explorations began with prototypes, including a chat-based system focused specifically on generating slide decks (PPT). This prototype was showcased at an internal demo event within our organization, where users of various professional backgrounds tried the system in an impromptu, in-situ manner. The sessions were not scheduled or guided, allowing us to observe how users naturally engaged with the interface. The simplicity of the chat interface, combined with its familiar resemblance to commercial generative AI systems like Microsoft Copilot and ChatGPT, facilitated a "walk up and use" strategy~\cite{olsen2007evaluating}. While this approach allowed for broad initial feedback, these early observations critically highlighted significant limitations in the prototype concerning transparency, user control, and the handling of diverse or multimodal content types.

We also observed that there is a wide diversity inherent in tasks such as document and slide generation, where user's expertise and approaches can vary significantly. We recognized that relying solely on traditional usability metrics might lead to a usability trap''\cite{olsen2007evaluating}, wherein standardized tasks for such a system are difficult to define meaningfully, and expectations of immediate proficiency are unreasonable given the complexity and variability of the work. Rather, we adopted an approach that involved rapid iteration and feedback from peers, complemented by dogfooding\cite{shneiderman2008creativity}. This approach proved to be an effective vehicle for understanding multifaceted problems within this design space and is loosely modeled after the thinking through prototyping'' suggested by Klemmer et al. \cite{klemmer2006bodies}.

Through iterative refinement,  we identified the need for structured yet flexible operations that users could transparently guide. We term these modular, user-controlled transformations \textit{infomorphs}, and use them as the fundamental unit for exploring this design space. Our system, in its various iterations, acted as a vehicle to unravel the different design dimensions of such infomorph-driven workflows.

This design space comprises five key dimensions derived from this process: (1) \textit{Perceptual Cues and Information Scent}, (2) \textit{User Control and Workflow Involvement}, (3) \textit{Interaction Granularity and Structural Agency}, (4) \textit{Generative Capabilities and Transformation Types}, and (5) \textit{Input \& Output Modalities}. 

For each of these dimensions, we highlight the primary guiding question that one would need to ask, in order to effectively plan the design, development and evaluation of Generative AI systems for document-based knowledge work. Although these dimensions and guiding questions may not be comprehensive, we believe that they will provide a structured starting point for systematically thinking, reasoning, and evaluating these systems. They may also help identify gaps for guiding future research and engineering efforts in this domain.

\subsection*{D1 - Perceptual Cues and Information Scent}
\label{dg:D1}
\vspace{0.2cm}

\hspace{0.25cm}\textbf{Q:}  ``\textsf{How effectively does the system enable users to anticipate the value of information before fully engaging with it?}''

\vspace{0.2cm}

Drawing directly from the Information Foraging Theory~\cite{ift95}, we see that the ability to \textit{quickly} and \textit{easily} assess the relevance of an information source, allows users to make informed decisions about which source is worth deeper exploration. Specifically, the concept of ``information scent''~\cite{10.1145/2600428.2609626,10.1145/365024.365331,informationscent} captures how users use perceptual cues to estimate the value of unseen content. Systems that offer strong perceptual cues reduce the cost of exploration and reduces iteration, by helping users efficiently prioritize their attention and effort, and helping choose the right set of sources. In the context of document transformation and generation, this emphasizes the importance of designing systems that make relevant content discoverable through \textit{intuitive} and \textit{informative visual and structural cues}. By enhancing information scent, such systems can not only help users in just choosing the right sources efficiently, but also allow users to better scope and define what kinds of infomorphs might be possible or useful. e.g., whether a source document is better suited for summarization, extraction, or translation. In this way, perceptual scent becomes a front-end affordance for surfacing candidate infomorphs early in the workflow.

To that end, we have the following sub-dimensions each of which may contribute to a richer information scent and perceptual cues.

\begin{itemize}
    \item[\textbf{D1.1}] \textbf{Surface-Level Previews:} Components such a brief excerpts, summaries, section outlines, thumbnails and any on hover interactions that offer immediate visual or textual cues about the underlying content. These support rapid scanning and help users identify contents, and parts of interest at a glance.
    \item[\textbf{D1.2}] \textbf{Source Metadata:} Information and metadata about the information source itself, such as document type, publication date, author name, number of pages, and associated tags or labels. These cues act as possible filters, helping users assess relevance, credibility, and scope before committing to deeper engagement.
    \item [\textbf{D1.3}] \textbf{Scent-optimzied Representations: } Signals that adapt to the perceived intent of the user, surfacing the content most likely to be useful in a given task. Some examples of this include keyword highlights, semantic tags, fast relevance-ranked suggestions and retrieval, or clustered excerpts\cite{Scim} drawn from multiple documents.  
\end{itemize}

\subsection*{D2 - User Agency and Workflow Guidance}
\label{dg:D2}
\vspace{0.2cm}

\hspace{0.25cm}\textbf{Q:}  ``\textsf{How actively does the user guide, interact with, and understand the generative document workflow?}''

\vspace{0.2cm}

Generative AI systems for document creation should support a broad range of user needs —from simple, automated outputs to richly customized, goal-driven workflows. Drawing on the ``low threshold, high ceiling, wide walls'' principle from creativity support tools articulated by Shneiderman \cite{shneiderman2008creativity}, effective systems must be easy to start with (low threshold), capable of supporting complex and expert-level tasks (high ceiling), and flexible enough to accommodate diverse workflows and goals (wide walls). Balancing automation with user control ensures that both casual users and power users can operate effectively within the same system.

This dimension captures how much control, and awareness a user can have in shaping how the system operates. This maybe through initiating actions, guiding and iterating intermediate steps, or even simply understanding what the system is doing and why. A system with strong support across this spectrum can allows users to first cold start with minimal input, and then progressively learn and use more control as needed. It supports exploratory workflows, but as an artifact, it can also enable iterative, transparent, feedback-driven refinement and co-creation with AI systems\cite{amershi2019guidelines}. We list some levels of this dimension here:

\begin{itemize}
    \item[\textbf{D2.1}] \textbf{Single/Few shot Generation:} Minimal input by the user, where the system outputs a complete draft without further user intervention.
    \item[\textbf{D2.2}] \textbf{Chain-of-thought / Prompt-driven Iteration:} Users refine outputs by adjusting prompts in multiple rounds.
    \item[\textbf{D2.3}] \textbf{Interactive Co-creation:} The system proposes intermediate content or outline options - users accept, reject, or refine them collaboratively.
    \item[\textbf{D2.4}] \textbf{Fully User-controlled Flow:} User defines each step of the generative process choosing which documents to include, specifying how content should be extracted, determining the transformation logic, and setting formatting or layout rules.
    \item[\textbf{D2.5}] \textbf{Feedback Loops (Cross-cutting):} This corss-cutting sub-dimension spans across others. Feedback mechanisms can be applied regardless of whether the user is operating at any of the above levels. It allows users to provide feedback such as inline edits, preference adjustments, user approval, highlight-to-improve, or that the system incorporates to adapt its behavior dynamically. Feedback loops have the potential to enhance awareness, responsiveness and conversational iteration at a granular level of the document creation process.   
\end{itemize}

 Each of these levels lists ways in which users can shape, inspect, and iterate on infomorphs. As workflows grow more complex, the need for modular, controllable infomorphs becomes critical to maintain transparency and adaptability.

\subsection*{D3 - Interaction Granularity and Structural Control}
\label{dg:D3}
\vspace{0.2cm}

\hspace{0.25cm}\textbf{Q:}  ``\textsf{At what level of detail can/ the user influence the form, structure, and composition of the generated content?}''

\vspace{0.2cm}

While D2 focused on the amount of detail at which a user is able to orchestrate, and understand how the system needs to translate from input sources to output a document of interest, there is another aspect of user involvement that centers not on control of the workflow, but on directly on the output generation itself.  This dimension at what level the user interacts with, and shapes the content and structure and has the following levels:

\begin{itemize}
    \item [\textbf{D3.1}] \textbf{Intent-level:} Users provide broad, purely, intention-level inputs. The system is responsible for inferring both structure and content with minimal user guidance. e.g. ``\textsf{Make a presentation about Nielsen's usability heuristics.}''
    \item [\textbf{D3.2}] \textbf{Format-level:} Users specify constraints or desired forms such as bullet points, tone, section count. They still delegate structural design, and content to the system. e.g. ``\textsf{Create a 5-slide deck explaining usability heuristics to HCI graduate students. Keep it academic and example-driven.}''
    \item [\textbf{D3.3}] \textbf{Collaborative Co-structuring:} The system proposes an initial structure such as section headers or an outline, which the user can accept, edit, reorder, or add/remove before content generation begins. E.g. The system proposes: \textit{(1) Definition and Origin; (2) Heuristics Overview (3) Applications in Practise; (4) Caveats and Challenges; (5) Moving Beyond. }
    In turn, the user modifies (2) and (4) as : ``\textsf{(2) Deep Dive: 10 Nielsen Principles; (4) Other Heuristic Models (e.g. Gerhardt-Powals)}'' keeping the rest unchanged. The system fills in the content, from the provided information sources using this revised structure.
      \item [\textbf{D3.4}] \textbf{User-defined structure:} The user defines a detailed outline, and\/or content wherever applicable, with explicit instructions (as required) for each section's role, scope, and data sources. The system executes content generation closely following these instructions. e.g. ``...\textsf{Slide 2: Overview of Heuristics in HCI - Image from \texttt{CS1337-Fall21-Lec42.pdf};...''}
      \item [\textbf{D3.5}] \textbf{Full Control:} The user has complete authorship over the document's structure, layout, and content. They have the ability to determine every part of the output, from the ordering of ideas and visual arrangement to individual slide content. The system in this case, serves as an assistant that fills in content only when requested, but every piece is editable, overrideable, and under direct user control.
\end{itemize}

It is important to note, these levels need not be exclusive. Ideally, a well-designed system would support these sequentially or contextually, allowing users to switch between them fluidly as desired, depending on their level of experience with the system, context of the task, and the quality of output by the system. For example, a user might begin with a simple request (D3.1), then move to revise an outline suggested by the system (D3.3), and finally make minor direct edits in the output themselves (D3.5). Such progressive control aligns with Shneiderman's ``low threshold, high ceiling, wide doors'' principle \cite{shneiderman2008creativity}, supporting both ease of entry and complex task expressivity.

\subsection*{D4 - Infomorphs}
\label{dg:D4}
\vspace{0.2cm}

\hspace{0.25cm}\textbf{Q:}  ``\textsf{What kinds of infomorphs a system can perform? What types of transformations can it apply to information sources to re-express them in new formats, structures, or modalities?}''

\vspace{0.2cm}

The previous dimensions centered on user agency and control with the overall flow and the generated content. An also important aspect is the types of generative transformations the system can perform, regardless of how much control the user exerts. Here, we focus on the system’s ability to reshape content in meaningful ways that support cognitive and creative work involved in the document-based knowledge work. We draw inspiration from prior work on techniques for navigating document clusters \cite{scatter_gather}, and identify three broad classes of infomorphs that reflect how people think about and apply changes to content: Scatter, Gather, and Transduce.

\vspace{0.2cm}
\noindent\textbf{D4.1 Scatter: }Scatter infomorphs break down, extract, filter, or reorganize information sources to surface meaningful parts or alternative framings, aiding exploration and sensemaking, particularly early in a workflow, where users are still making sense of an information source. With document-based knowledge work, such infomorphs can particularly be helpful for users in helping unpack large pieces of unstructured information that maybe interleaved with text, images, charts etc,. . Rather than synthesizing polished outputs, scatter capabilities aim to reveal what’s inside an information source, and how it might be reapproached or repurposed. Some examples include:
\begin{itemize}
    \item Extracting relevant sections or pages from a long report
    \item Surfacing all figures, tables, or annotated highlights from a PDF \cite{Scim}
    \item Extracting raw information clusters for a downstream processing task
    \item Performing filters on information source metadata, such as author, time, etc,.
\end{itemize}

\vspace{0.2cm}
\noindent \textbf{D4.2 Gather: }While scatter infomorphs help unpack and explore information by breaking it into meaningful parts, Gather infomorphs enable the opposite direction.  They help bring together different pieces into more coherent, structured forms, for the purpose of synthesizing new narratives drawn from different fragments of information sources. In practise, for document-based knowledge work, their output serve as intermediate artifacts, that are open to further refinement. While these outputs can sometimes serve as final deliverables, they may frequently lead into another set of infomorphs - \textit{scatter}, \textit{gather}, or \textit{transduce} infomorphs, reflecting the iterative and recursive nature of complex knowledge tasks. Some examples include:
\begin{itemize}
    \item Merging relevant information from papers into a unified slide deck for a spefic purpose
    \item Integrating content across modalities (text, visuals, annotations)
     \item Assessing and assembling related information from diverse sources into a unified structure.
\end{itemize}

\vspace{0.2cm}
\noindent \textbf{D4.3 Transduce: }Transduce infomorphs re-expresses information, such as changing its form, style, or modality, while majorly preserving the underlying content and structure. These operations are typically used during the later stages of a workflow, when the focus shifts from exploration or synthesis to preparing content for a specific audience, format, or communication medium. Unlike scatter and gather, which reshape the organization or composition of content, transduce focuses on the form in which a content is presented and delivered.

An \textbf{infomorph} is not a monolithic transformation, but a modular unit that can be composed with others. For example:
\begin{itemize}
    \item \textit{Extract top quotes from interview transcripts} (\texttt{scatter})
    \item \textit{Merge quotes into thematic categories} (\texttt{gather})
    \item \textit{Reformat themes into a visual timeline} (\texttt{transduce})
\end{itemize}

These transformation types are not mutually exclusive. In practice, workflows often chain together multiple \textit{infomorphs}. Such chaining of infomorphs allows users to iteratively build structured, multimodal documents while retaining user agency and control. This compositionality is key to complex knowledge work.

\subsection*{D5 - Input and Output Modality}
\label{dg:D5}
\vspace{0.2cm}

\hspace{0.25cm}\textbf{Q:}  ``\textsf{What types of input and output formats or modalities are supported by the generative system?}''

\vspace{0.2cm}

Effective document-based knowledge work often involves engaging with diverse types of content, both as input sources and as target output content. On both the input and output sides, systems may support a range of content types and modalities, including unstructured or structured text such as markdown or form-based inputs, word documents, powerpoints, visual materials like images, diagrams, and multimodal documents that combine formats, such as PDFs or webpages. Some systems may also process audio or video content, including transcripts or recorded material. However, support for handling multiple modalities within a single system remains limited in many current tools.
\\
\\
\noindent \textbf{Summary}
\noindent There are many Generative AI tools for Document creation today, and their number is increasing every day. This makes it challenging to understand and characterize their underlying capabilities and interaction patterns. To help navigate this complexity, we propose analyzing these tools through a design space defined by five key dimensions, encompassing aspects like information scent, user agency, interaction granularity, transformation types, and supported modalities. Our exploration within this design space highlighted the need for a framework centered on \textit{infomorphs}, modular operations designed for information transformation. 
As a specific instantiation within this design space, in the next section, we introduce \systemnamedot, which uses a canvas-based interface for composing infomorphs; though it is not necessarily the only or optimal approach for realizing these principles. Regardless of the specific interface, the core principles aim to better support the complex, iterative, and multi-modal demands of modern knowledge work.
\section{\systemname - Scenario Walkthrough}

\systemname is a canvas-based system for building interactive, multimodal document transformation, and generation using infomorphs. It supports complex document-based knowledge work by enabling users to \textit{scatter}, \textit{gather}, and \textit{transduce} information across a wide range of formats and sources. The system combines a visual workflow canvas with a conversational interface (\textit{chat}). Users can upload content such as PDFs, Word docs, PPTs, spreadsheets, and webpage links, then compose \textit{infomorphs} that can transform the content in these sources, into desired outputs. Each \textit{infomorph} can be used to guide, and refine multimodal information to reflect the user's specific goals. Throughout the process, intermediate results remain visible and editable, allowing users to explore options, provide feedback, and iteratively shape their outputs. This supports transparent, modular, and intent-driven workflows for tasks such as planning, summarization, and presentation authoring.

To illustrate \systemnamedot’s capabilities in a realistic workflow, we present a usage scenario following Alex, an HCI researcher attending UIST 2025 in Busan (September 28 to October 1, with trip dates from September 26 to October 3). To ensure ecological validity, we curated content from real sources, adapting the 2024 UIST program to the 2025 calendar and incorporating leisure activities from actual travel guides with some creative flexibility.

This task represents complex knowledge work involving diverse information sources, planning, budgeting, iteration, and generating multiple outputs (itinerary, budget, and a summary presentation), highlighting the system's capability in multimodal synthesis, particularly slide deck generation, and iterative updates.
\vspace{-5pt}

\subsection*{Chat-based Initiation}

\begin{figure}[h] 
\centering 
\fbox{\includegraphics[width=\linewidth]{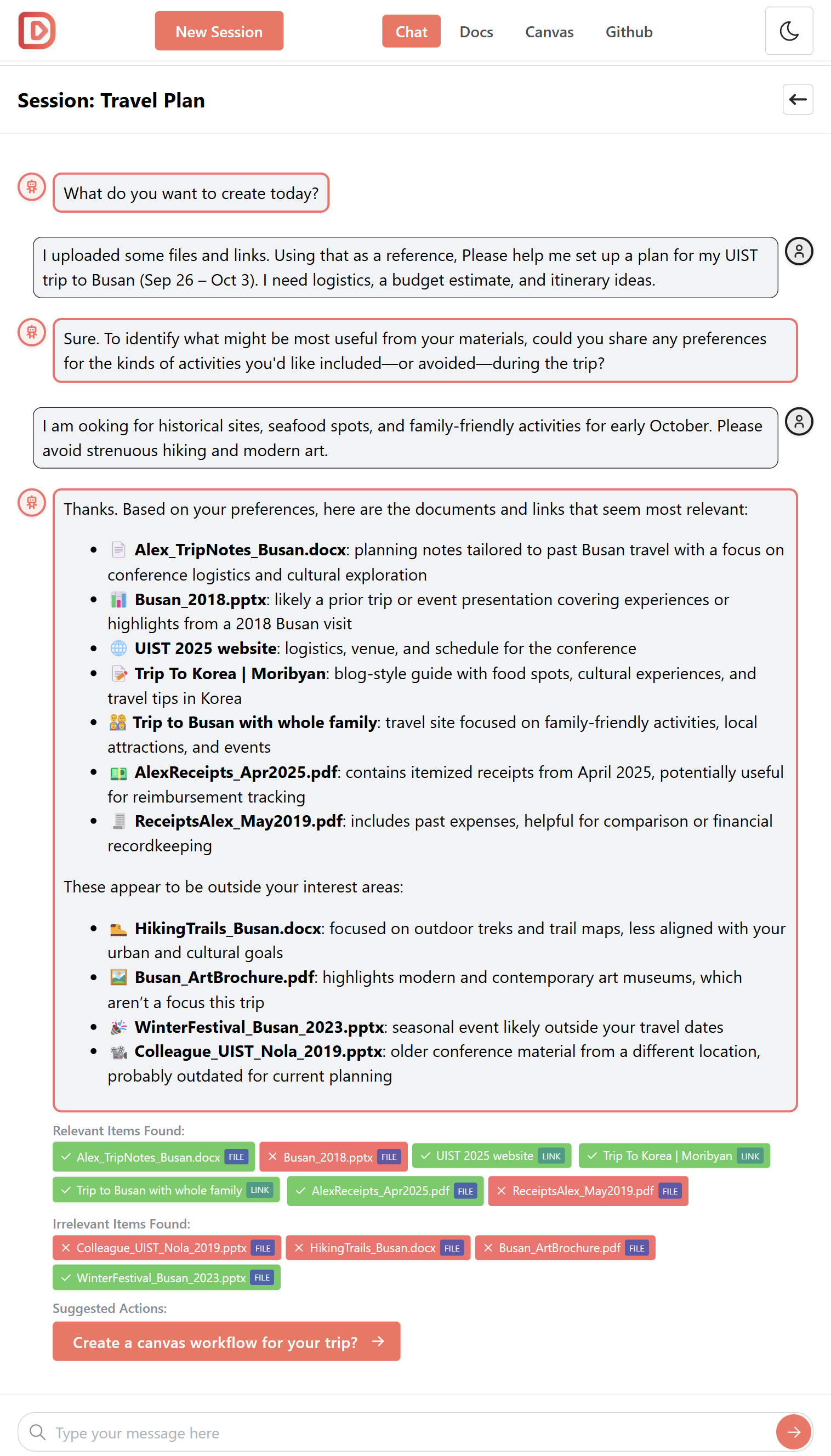}} 

\caption{Initial chat interaction in \systemnamedot. After Alex uploads files and pastes links, the system prompts for activity preferences and returns a list of relevant and irrelevant information sources. This supports source-level triage based on user intent, illustrating information scent (D1), preference-aware prompting (D2.2), and scatter infomorph behavior (D4.1).} 

\end{figure} 

Alex begins by opening \systemname’s chat interface and uploads a range of input sources (\hyperref[dg:D5]{\textbf{D5i}}), that he had collected in a folder from a quick skim of the internet and his filesystem. Alex types an initial planning prompt:

\vspace{2pt}
 
\texttt{``Please help me set up a plan for my UIST trip to Busan (Sep 26 – Oct 3). I need logistics, a budget estimate, and itinerary ideas.''}

\vspace{4pt}

The system follows up with a clarification request, asking Alex to specify interests and constraints (\hyperref[dg:D2]{\textbf{D2.2}}, \hyperref[dg:D3]{\textbf{D3.1}}). Alex responds: 

\vspace{4pt}

``\texttt{Looking for historical sites, seafood spots, and family-friendly activities for early October. Please avoid strenuous hiking and modern art.}''  

\vspace{4pt}

Following that, the system analyzes the content of the documents and links uploaded, presenting a categorized list of sources based on inferred relevance. Items such as travel blogs, trip notes, and flight confirmations are identified as useful, while materials such as the hiking guide and art brochure are flagged as unrelated based on Alex’s preferences. He notices that two of the relevant files (\filename{Busan\_2018.pptx} and  \filename{ReceiptsAlex\_May2019.pdf}) maybe outdated, while (\filename{WinterFestival\_Busan\_2023.pptx}) might be relevant. So, he directly modifies the relevance through the chat.

This interaction highlights the system’s early-stage support for perdicted user intent, and information scent (\hyperref[dg:D1]{\textbf{D1.1}}), where users are able to quickly identify which documents are likely to be valuable. It also showcases the use of prompt-driven iterations (\hyperref[dg:D2]{\textbf{D2.2}}) and reflects the behavior of \textit{scatter} infomorphs at the document level (\hyperref[dg:D4]{\textbf{D4.1}}), potentially setting the ground for further information processing. 

At the end of the chat interaction, the system offers to continue in the canvas view:
\texttt{``Create a canvas workflow for your trip?''} and Alex proceeds to the Canvas

\vspace{-5pt}

\subsection*{Canvas-based \textit{infomorph} workflow}

\begin{figure*}[hbt!] 
\centering 
\includegraphics[width=\linewidth]{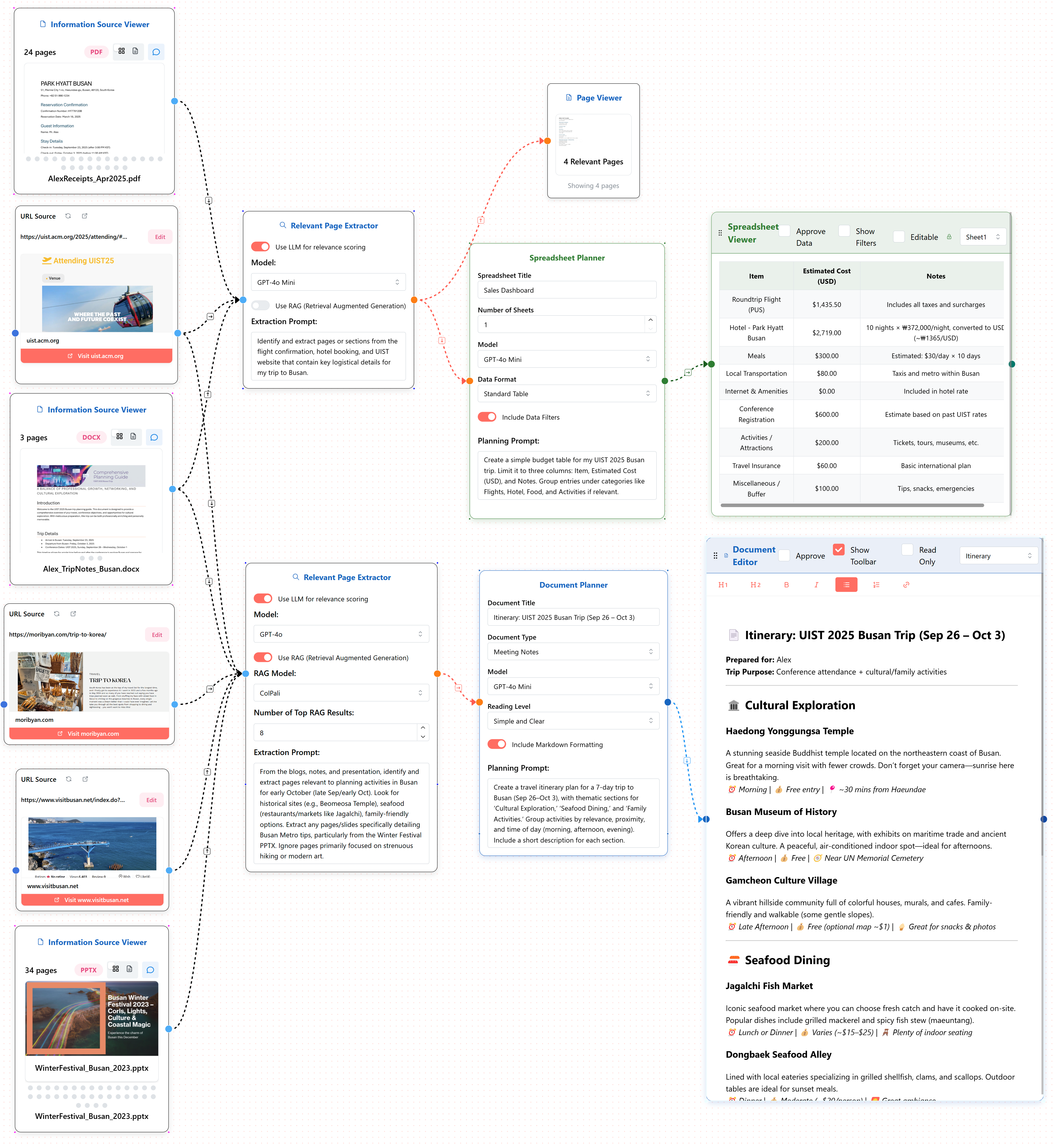}

\caption{Initial canvas view auto generated by \systemname based on Alex's planning chat conversation \hyperref[dg:D2]{D2.2} $\rightarrow$ \hyperref[dg:D2]{D2.4}. The workflow splits into two parallel branches: (top) logistics and budgeting, and (bottom) itinerary planning. Each branch chains together modular \textit{infomorphs}, such as \imnode{Relevant Page Extractor}, \imnode{Planners}, \imnode{Viewers} and \imnode{Editors}, allowing Alex to inspect, refine, and generate structured outputs. The canvas illustrates a transparent, human-in-the-loop transformation process across multiple modalities (PDFs, URLs, PPTs, DOCs, and XLs).} 
\label{fig:initial_flow}
\end{figure*}

Using the earlier conversation and the refined list of sources, \systemname algorithmically generates an initial flow composition of \textit{infomorphs} that Alex can inspect and iterate on. This marks a shift from prompt-driven interaction (\hyperref[dg:D2]{\textbf{D2.2}}) to a structured, user-steerable workflow  (\hyperref[dg:D2]{\textbf{D2.4}}), illustrating how \systemname fluidly bridges conversation and information flow composition.

Broadly, the canvas displays the selected input sources (Word docs, PDFs, PPTX files, web links via \imnode{Source Nodes}) feeding into two main parallel branches, reflecting Alex's request for logistics/budget and itinerary ideas (\hyperref[dg:D5]{\textbf{D5i}}), and is shown in the Figure \ref{fig:initial_flow}
\\
\\
\scenarioheading{Branch 1: Logistics and Budget}

Relevant sources (\filename{Alex\_TripNotes\_Busan.docx}, \filename{UIST' 25 website}, \filename{AlexReceipts\_Apr2025.pdf}) are connected to a \imnode{Relevant Page Extractor} node (\hyperref[dg:D4]{\textbf{D4.1 Scatter}}), configured to pull out pages containing details on flights, accommodation hints, conference dates, and known expenses. Since such logistics information is critical, \imnode{Relevant Page Extractor} ensures that each page of the source is ``looked at'' by an LLM for retrieval. While this adds to latency of the system, it also improves accuracy. The system also adds a \imnode{Page Preview} node so that Alex can view the actual source pages that were used for information generation, enhancing transparency (\hyperref[dg:D1]{\textbf{D1.1}}, \hyperref[dg:D1]{\textbf{D1.3}}). 

The output of the extractor flows into a \imnode{Spreadsheet Planner} (\hyperref[dg:D4]{\textbf{D4.2 Gather}}) node. Drawing from Alex's initial chat request for a budget draft, the system prepopulates the planning prompt field in the node's configuration panel (Fig. \ref{fig:initial_flow}) with: 

\vspace{4pt}

\texttt{``Create a simple budget table for my UIST 2025 Busan trip. Limit it to three columns: Item, Estimated Cost (USD), and Notes. Group entries under categories like Flights, Hotel, Food, and Activities if relevant.''}

\vspace{4pt}

Alex reviews this system-suggested prompt. To ensure the estimations are clearly justified in the output, he decides to modify the prompt slightly by adding: 

\vspace{4pt}

``\texttt{In the Notes column, please specify the source or assumption for each estimate where possible.''} 

\vspace{4pt}

\begin{figure*}[htb] 
\centering 
\includegraphics[width=\linewidth]{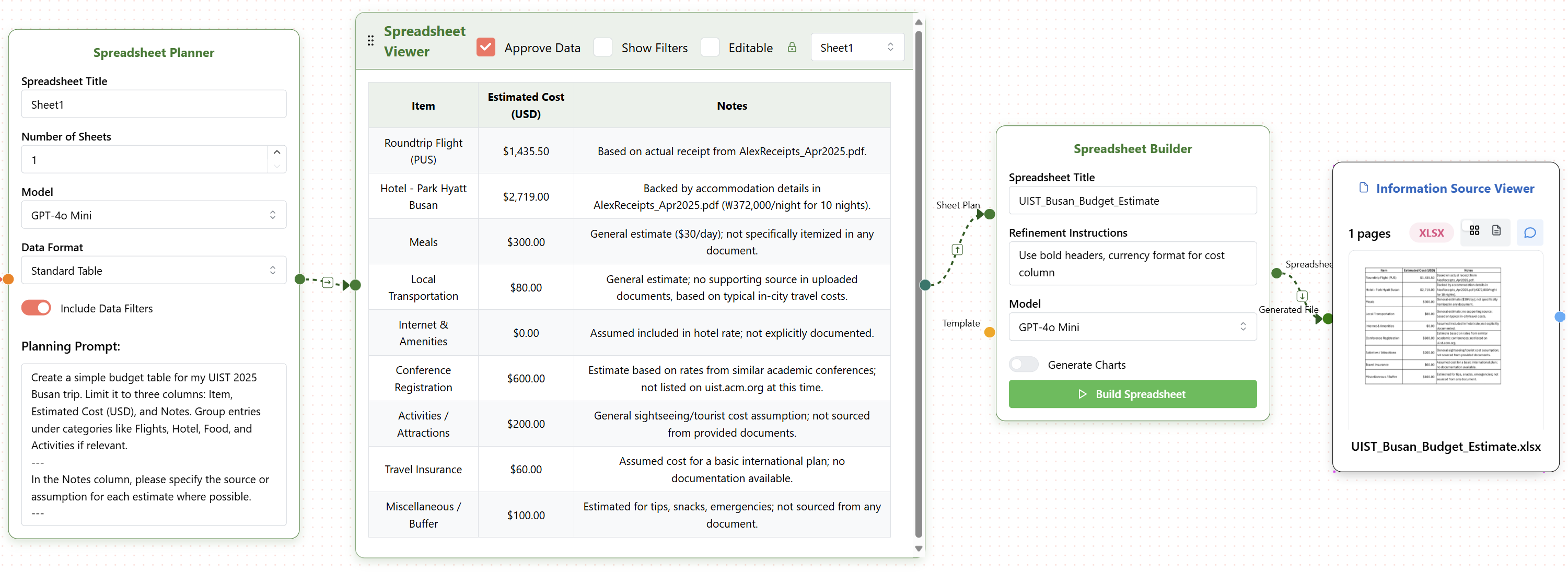}

\caption{Review and export of the budget estimate co-created by \systemname and Alex as formatted XLSX file.} 
\label{fig:export_estimate}
\end{figure*}

This step demonstrates Alex steering the generation by refining the instructions provided to the AI (\hyperref[dg:D2]{\textbf{D2.3 Interactive Co-creation}}, \hyperref[dg:D3]{\textbf{D3.3 Collaborative Co-structuring}}). The rest of the node settings, such as the AI model (GPT-4o Mini), data format (Standard Table), and data filters are kept as pre-configured (\hyperref[dg:D3]{\textbf{D3.2 Format-level}} guidance via prompt structure) (Figure \ref{fig:export_estimate}).
\\
The instructions from the Spreadsheet Planner guide the generation of the budget content, which is then displayed directly in a Spreadsheet Viewer (\hyperref[dg:D5]{\textbf{D5o}}) node (see Figure \ref{fig:export_estimate}). This viewer presents the interactive table with ``Item'', ``Estimated Cost (USD)'', and ``Notes'' columns, populated according to Alex's refined instructions. The ``Notes'' column reflects his request for sources/assumptions. The viewer includes controls like "Approve Data", "Show Filters", and "Editable", allowing for validation and direct interaction (\hyperref[dg:D2]{\bf{D2.5 Feedback Loops}}, potentially \hyperref[dg:D3]{\textbf{D3.5 Full Control}} via editing).  
\\
\\
\scenarioheading{Branch 2: Itineary Ideas}

Some of the relevant sources identified earlier, \filename{Alex\_TripNotes\_ Busan.docx}, \filename{moribyan.com},\filename{visitbusan.net}, \filename{UIST'25 website}, and \filename{WinterFestival\_Busan\_2023.pptx} are connected to the \imnode{Relevant Page Extractor} node (\hyperref[dg:D4]{\textbf{D4.1 Scatter}}), based on the overall task context and Alex's preferences gathered from the initial chat (\hyperref[dg:D1]{\textbf{D1.3}}). 
\\
\\
As shown in its configuration panel (see Figure\ref{fig:initial_flow}), this \imnode{Relevant Page Extractor} node uses a two-stage approach for efficiency. First, a fast Retrieval-Augmented Generation (RAG) model (ColPali \cite{colpali}) performs an initial pass to retrieve the `Top k' (k=8) most relevant pages, or content chunks (for URLs). Second, a multimodal LLM (GPT-4o) `looks at' each of these top 8 RAG results. This offers a good balance between latency, and accuracy. This model utilizes the detailed `Extraction Prompt' which the system also pre-populated based on the preferences Alex stated during the initial chat (e.g., \textit{Alex: }"\textit{...specifically focus on historical sites, seafood dining, and family-friendly activities... Avoid strenuous hiking and modern art content...}") (\hyperref[dg:D1]{\textbf{D1.3}}). Alex reviews this pre-filled extraction prompt within the node's configuration panel, and confirms it accurately captures his requirements before proceeding (\hyperref[dg:D2]{\textbf{D2.3}} through reviewing system suggestions). This ensures faster responses, while also not corrupting the context of the LLM with content that is not very related to Alex's intent (\hyperref[dg:D4]{\textbf{D4.1 Scatter}}). Similar to the counterpart in Branch 1, the relevant pages and chunks extracted flow, now into the \imnode{Document Planner} (\hyperref[dg:D4]{\textbf{D4.2 Gather}}). Similar to the extractor, the system pre-populates the planning prompt here based on the goal of creating an itinerary. Alex reviews this prompt ("\textit{Create a travel itineary covering the 7-day trip... each section}") and potentially refines it to adjust the structure or emphasis (\hyperref[dg:D3]{\textbf{D3.2},\textbf{D3.3}}). This planning stage uses a separate model (GPT-4o Mini). The final itineary content, generated according to the Document Planner's instructions, is displayed in the Document Editor (\hyperref[dg:D5]{\textbf{D5o}}). The editor shows the draft itinerary, organized thematically as requested. Alex can review the content, use controls like ``Approve'', or make direct inline edits to the text (\hyperref[dg:D3]{\textbf{D3.5}}).

\vspace{3em}

\scenarioheading{Approving Initial Plan Components}

Before the conference details are finalized, Alex reviews the budget displayed in the interactive Spreadsheet Viewer and the thematic itinerary draft in the Document Editor. Happy with these outputs resulting from his collaboration with the system, he clicks the ``Approve'' button available on both nodes (\hyperref[dg:D2]{\textbf{D2.5 Feedback Loops}}). This signifies his acceptance of the current state and locks the content within these nodes. Now, even if upstream data sources were altered, these approved nodes would retain their current content unless Alex explicitly chose to unapprove and recompute them, preserving the specific versions he co-created with \systemname (\hyperref[dg:D2]{\textbf{D2.3}}, \hyperref[dg:D3]{\textbf{D3.5}}).

With the initial budget draft reviewed and approved in the Spreadsheet Viewer, Alex needs to submit this estimate to his institution for travel pre-approval. This step is crucial before finalizing bookings. He adds a \imnode{Spreadsheet Builder} node (\hyperref[dg:D4]{D4.3 Transduce}) onto the canvas, connecting the output of the approved Spreadsheet Viewer node to it. Triggering the Spreadsheet Builder takes the current state of the approved data, along with some formatting instructions that Alex enters, and generates the final \filename{UIST\_Busan\_Budget\_Estimate.xlsx} file (Figure \ref{fig:export_estimate}). Alex downloads this file from the resulting Document Source node (\hyperref[dg:D5]{D5o}), ready for his institutional submission process. This completes a key administrative step well before the trip begins.

\scenarioheading{Integrating the Final Conference Schedule}

Weeks later, the finalized UIST 2025 program PDF, \filename{UIST\_ Program\_Final.pdf} becomes available. Alex uploads this new document to \systemnamedot's canvas. His goal is to merge the specific conference sessions he's interested in with minimal disruptive changes to his existing, approved itinerary.Since the itinerary is already in the \imnode{Document Editor} node and marked as ``Approved'' it remains frozen—preserving the co-created timeless itinerary from earlier stages.

\noindent \textbf{Extract Relevant session}: Alex adds a new \imnode{Relevant Page Extractor} (Scatter Infomorph – \hyperref[dg:D4]{\textbf{D4.1}}) to the canvas, linking it to the uploaded \filename{UIST\_Program\_Final.pdf}. He configures the node with the extraction prompt:

\vspace{4pt}

\texttt{``From the UIST program PDF, extract pages having talks to Human-AI Interaction.''}

\vspace{4pt}

\noindent \textbf{Merge and Re-plan}: To ensure he can attend some of these extracted talks, Alex connects the output of the extractor and the previously approved \imnode{Document Editor} node into a new \imnode{Document Planner} (\hyperref[dg:D4]{\textbf{4.2 Gather}}) with a short prompt: 

\vspace{4pt}

\texttt{``Merge the extracted UIST talks into the existing itinerary between Sep 28–Oct 1. Keep the structure intact, adjusting nearby items (like meals or free time) only when needed.''} 

\vspace{4pt}

\begin{figure*}[htb] 
\centering 
\includegraphics[width=\linewidth]{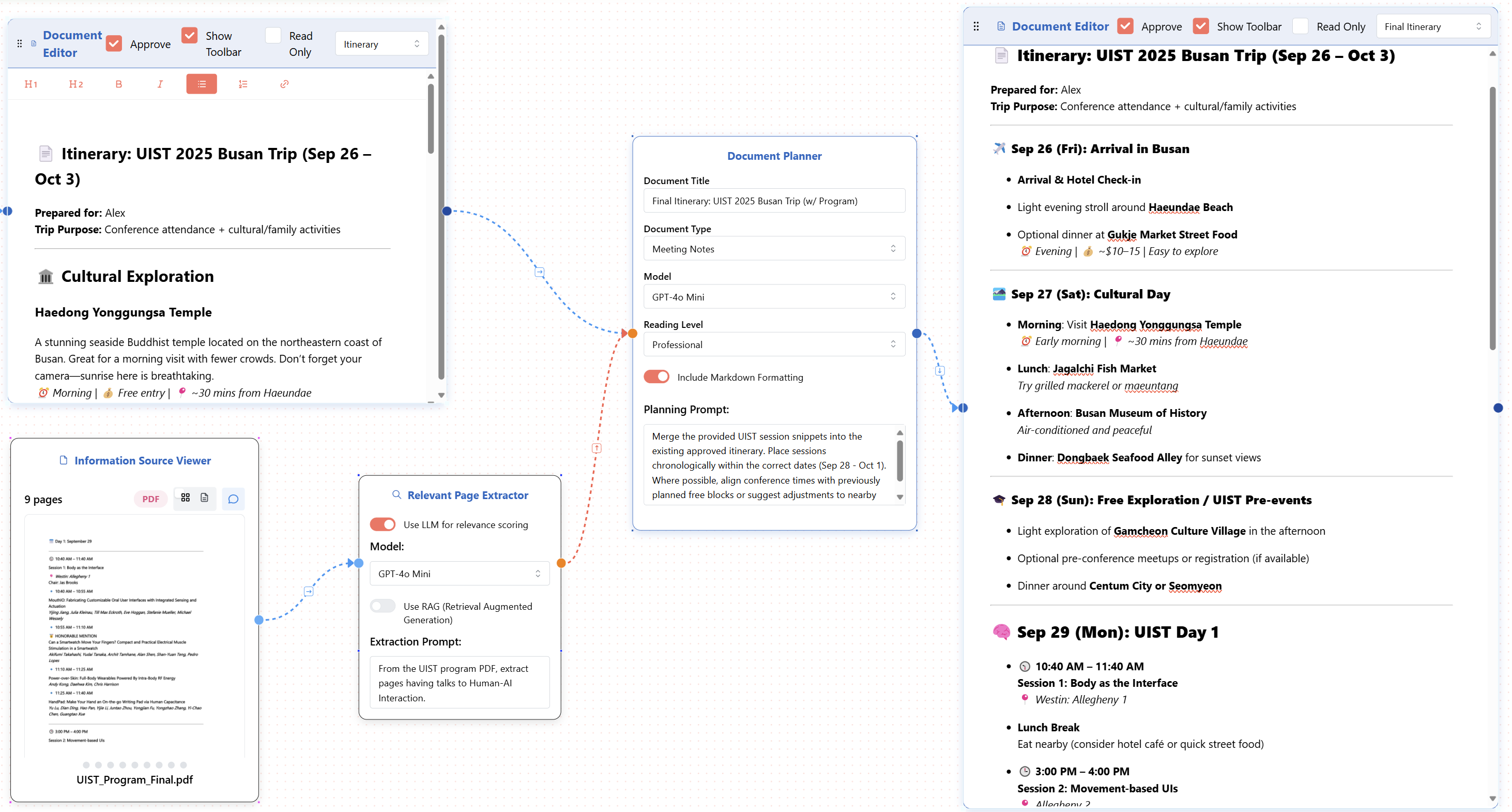}

\caption{Creation of revised itinerary by merging information from the previous itinerary and \filename{UIST\_Program\_Final.pdf}} 
\label{fig:revised_itinery}
\end{figure*} 

Alex then feeds the resulting merged plan into a new \imnode{Document Editor}. Upon executing the flow, He sees his planned visits to Gamcheon Culture Village and Jagalchi Market now interleaved with pre-conference meetups. He might make a few final manual tweaks for readability or to resolve minor timing conflicts (\hyperref[dg:D3]{\textbf{3.5 Full Control}}) before hitting approve. The results and the entire flow can be seen in Figure \ref{fig:revised_itinery}

\scenarioheading{Post-Conference: Generating a Summary Presentation}

After returning from Busan, Alex wants to synthesize his conference experience. Instead of taking notes in separate files during the event, he had conveniently used Document Editor nodes directly on his \systemname canvas each day (creating, for instance, \imnode{Notes\_Day1\_Editor}, \imnode{Notes\_Day2\_Editor}, and \imnode{Notes\_Day3\_Editor}). He captured his thoughts, key points from talks, as well as some images of the work within these nodes during the sessions. Now, back in the office, he aims to transform these canvas-based notes into an informative presentation for his colleagues who weren't there. Beyond using these notes, Alex also had to use the finalized travel itinerary in its Document Editor, to accurately reconstruct when and where each session occurred. Just to ensure all data is there, he also re-uses and the \textit{cached} results from the \imnode{Relevant Page Extractor} that originally served as a reference to the final itinerary. This helps fill in missing / incorrect speaker names, titles, and affiliations that weren’t captured in the session notes that Alex had taken. 

\begin{figure*}[htb] 
\centering 
\includegraphics[width=\linewidth]{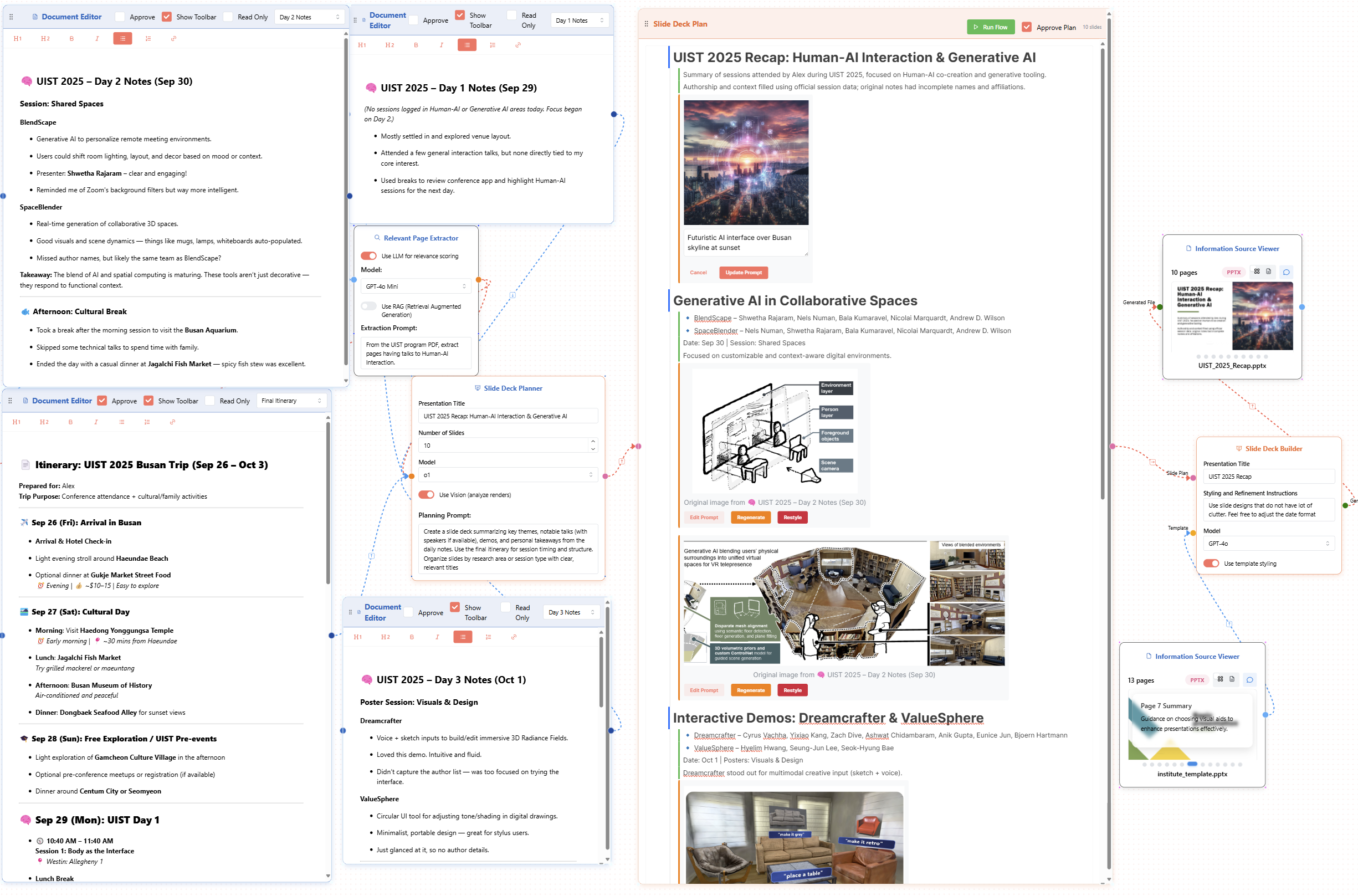}

\caption{The \systemname canvas during post-conference presentation synthesis. Daily notes and the final itinerary (from \imnode{Document Editor} and \imnode{Relevant Page Extractor} nodes) feed into the \imnode{Slide Deck Planner}. The resulting draft presentation is displayed and refined in the interactive \imnode{Slide Deck Plan Viewer} (center-right), showcasing AI-generated title imagery and content slides. These refined slides are subsequently exported using a \imnode{Slide Deck Builder} node which applies a designated institutional template.} 
\label{fig:generated_ppt}
\end{figure*}

He adds a \imnode{Slide Deck Planner} node, which uses these nodes as information sources, and provide the prompt:

\vspace{4pt}

\texttt{``Create a slide deck summarizing key themes, notable talks (with speakers if available), demos, and personal takeaways from the daily notes. Use the final itinerary for session timing and structure. Organize slides by research area or session type with clear, relevant titles''}

\vspace{4pt}

The \imnode{Slide Deck Planner} processes the gathered content, populating an interactive Slide Deck Viewer with the draft presentation (Seen in Figure \ref{fig:generated_ppt}) Alex makes a few edits here: He reviews and edits the generated text on slides, reorganizes content blocks. He observes that the initial title slide ("UIST 2025 Recap...") is purely text. To add to the visual appeal, he clicks the ``Generate Image'' button within the title slide's area and provides a prompt  to generate a new image. On subsequent slides, some images were pulled from his notes. For such, he might use the ``Restyle Image'' button to request stylistic changes to the image. This ability to restyle both newly generated and existing original images are instance of multimodal adaptations to previously mentioned design dimensions. (\hyperref[dg:D5]{D5o}, \hyperref[dg:D4]{D4.3 Transduce}). After iterating on text, structure, and visuals across all slides, he approves the final state within the viewer (\hyperref[dg:D2]{2.5}). 

To get the final, institutionally formatted PowerPoint file, Alex adds a \imnode{Slide Deck Builder} node . He connects the output of the approved \imnode{Slide Deck Viewer} to this builder. He configures the builder to apply his institution's presentation template (\filename{institution\_template.pptx}) (\hyperref[dg:D4]{\textbf{D4.3}}). Triggering the builder generates the final file .pptx file.

\section{Implementation}

\systemname is implemented using a client-server architecture designed to support the interactive, multimodal, and computationally intensive nature of \textit{infomorph}-driven document workflows.

\textbf{Frontend}: The user interface, including the interactive canvas, node configuration panels, chat interface, and content viewers/editors, is built as a web application using Mantine(React). We leverage the \textit{xyflow} library for rendering and managing the node-based graph visualization on the canvas. For rich text editing capabilities within specific nodes, we integrate libraries such as Tiptap and blocknote. The frontend communicates with the backend via REST APIs to trigger workflow operations, fetch results, and manage state.

\textbf{Backend}: The backend consists of a combination of Node.js and Python Flask services. It is responsible for orchestrating the execution of infomorph workflows, managing document parsing and processing, interfacing with various Generative AI models (both local and cloud-based), handling data persistence, and managing the computation graph.

\begin{figure*}[ht]
    \centering
    \begin{minipage}[t]{0.22\textwidth}
        \centering
        \includegraphics[width=\linewidth]{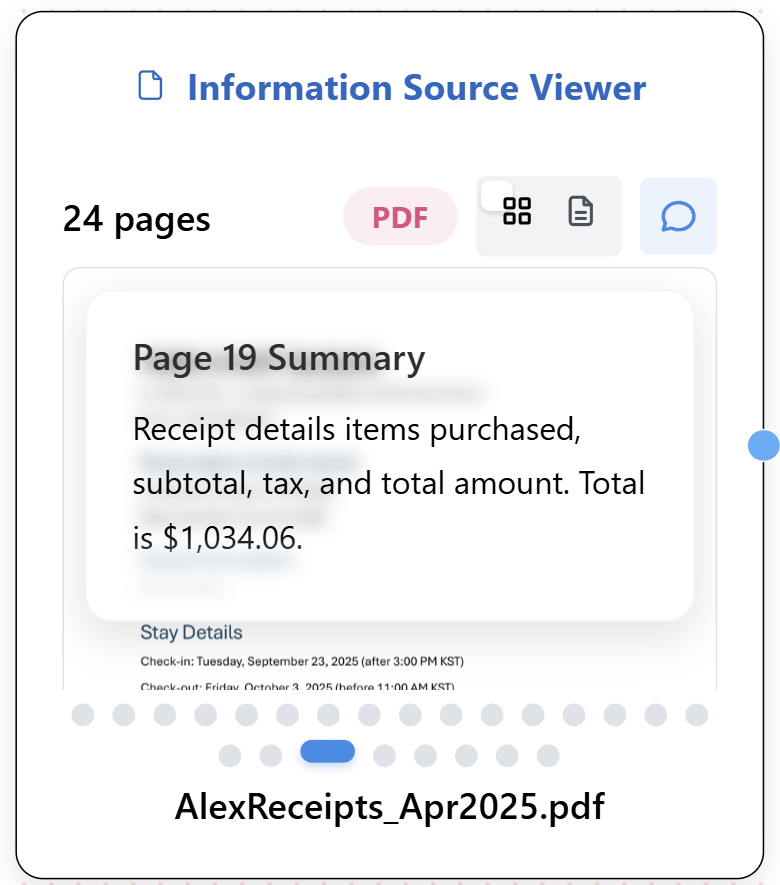}
        \caption*{}
    \end{minipage}
    \hfill
    \begin{minipage}[t]{0.22\textwidth}
        \centering
        \includegraphics[width=\linewidth]{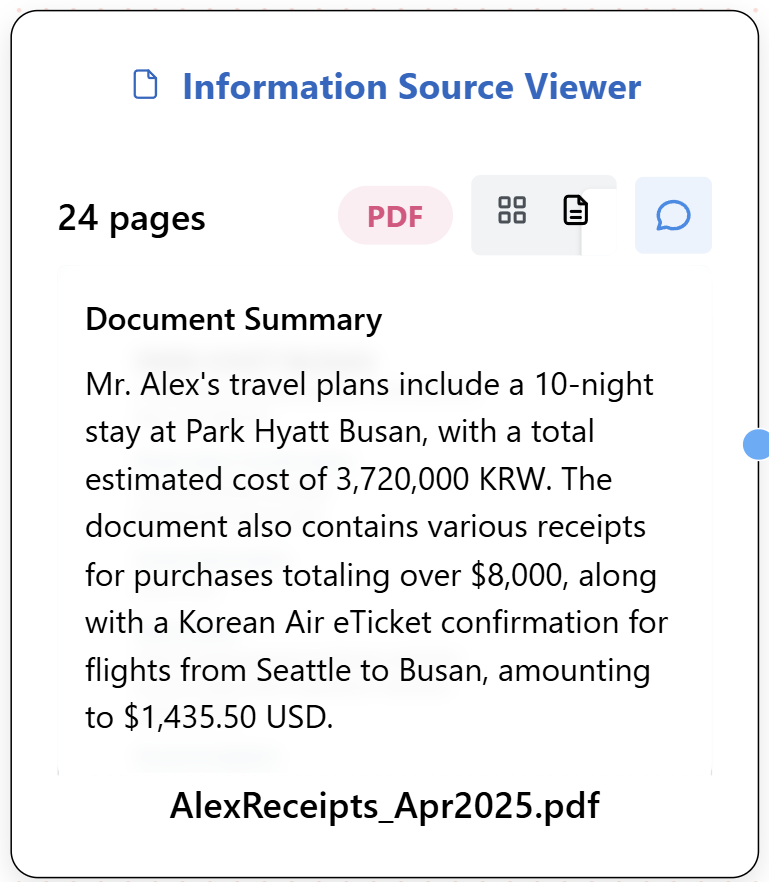}
        \caption*{}
    \end{minipage}
    \hfill
    \begin{minipage}[t]{0.4\textwidth}
        \centering
        \includegraphics[width=\linewidth]{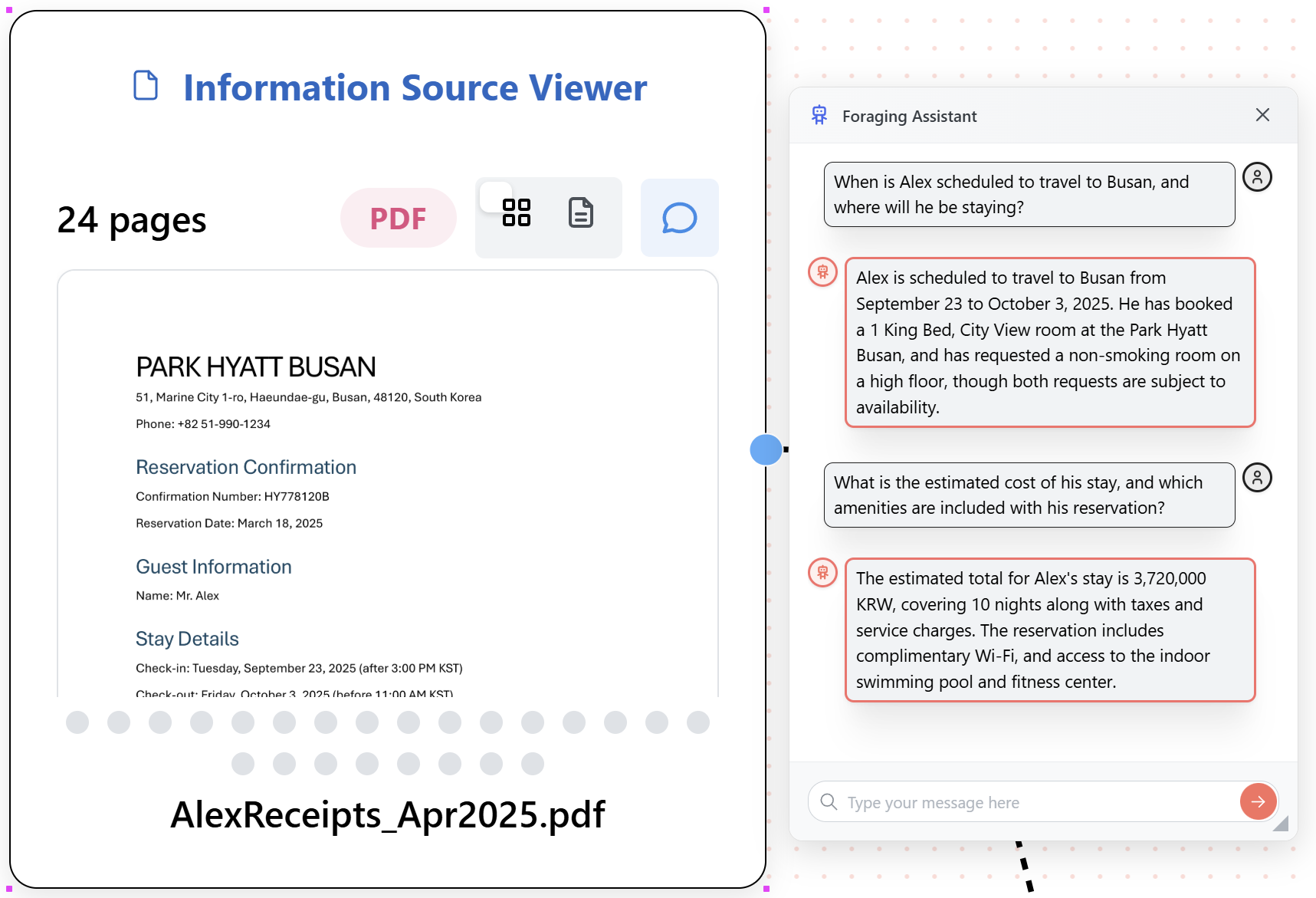}
        \caption*{}
    \end{minipage}
    \caption{Illustrating Information Scent (D1) features within the \texttt{SystemName} Information Source Viewer for a document (\texttt{AlexReceipts\_Apr2025.pdf}). (Left) Hovering over a page marker shows AI-generated summary for that specific page, enabling quick content checks. (Middle) Summary of the entire document. (Right) A chat interface scoped specifically to the file, facilitating interactive exploration of relevance and details.}
    \label{fig:three_minipage}
\end{figure*}

\subsection{infomorph  Implementation Details}

\imnode{File Source Node}, \imnode{URL Source Node}: These nodes represent the entry points for user-provided content (e.g., PDFs, DOCX, PPTX, URLs, spreadsheets). They handle the initial ingestion and parsing of different file types. To enhance information scent (\textbf{D1}), \imnode{File Source Nodes} offers several preview mechanisms visible on the node itself: hovering over page markers shows page summaries that were computed during document upload; users can click the toggle on the top right to see a summary for the entire document. Furthermore, users can initiate a dedicated, sandboxed chat conversation scoped specifically to converse about the contents of that file, allowing focused querying without context corruption. These different mechanisms are seen in Figure \ref{fig:three_minipage}. Rendering previews for Office documents (.docx, .pptx, .xlsx) is handled programmatically, often leveraging Microsoft Office APIs on the server side. When user clicks on a page, they can see it in full screen, to have a closer look. They can also resize the nodes in the canvas.

\imnode{Relevant Page Extractor}: This node embodies a Scatter infomorph (\textbf{D4.1}), designed to extract specific pages (from files) or content chunks (from url/html) based on relevance. As described in the scenario, it often employs a multi-stage process for efficiency: an initial fast retrieval stage using Retrieval-Augmented Generation (RAG) with vector embeddings identifies candidate chunks, followed by a more detailed analysis using a multimodal LLM (like GPT-4o) to refine the selection based on the user's extraction prompt and visual content if applicable. The backend supports various embedding models for the RAG stage, including text-only (e.g., OpenAI ADA embeddings), image-only (e.g., ColPali), and multimodal text+image embeddings (e.g., CLIP), chosen based on the content type and task requirements, selectable in the configuration panel. 

\textbf{Planning Nodes}  - \imnode{Slide Deck Planner}, \imnode{Spreadsheet Planner}, \imnode{Document Planner}: These nodes typically function as Gather infomorphs (\textbf{D4.2}). They take structured or unstructured data from upstream nodes (e.g., extracted content) and user-provided prompts as input . The backend service associated with these nodes invokes a configured LLM (e.g., GPT-4o Mini, GPT-4o) with the combined context (input data + prompt) to generate draft content structured according to the target format (e.g., slide outlines, table data, structured document sections).

\textbf{Building/Export Nodes} - \imnode{Slide Deck Builder}, \imnode{Spreadsheet Builder}, \imnode{Document Builder}: These nodes function as Transduce infomorphs (\textbf{D4.3}), responsible for converting the potentially edited and approved intermediate representations into final, standard file formats (e.g., .pptx, .xlsx, .docx). They often involve applying formatting rules or templates (e.g., using a provided organization .pptx template). 

\textbf{Viewing/Editing Nodes }-  \imnode{Slide Deck Plan Viewer}, \imnode{Spreadsheet Viewer}, \imnode{Document Editor} : These nodes present the intermediate outputs from Planner nodes in an interactive format on the frontend, allowing for inspection and refinement (Figure 10 provides an example). The \imnode{Document Editor} utilizes Tiptap for rich text interaction. The \imnode{Spreadsheet Viewer} employs a custom interactive data grid component. The \imnode{Slide Deck Plan Viewer} renders an interactive outline/preview, using blocknote for its text editing capabilities, where users can reorder items and trigger image operations via backend calls: fresh image generation is handled using models like DALL-E, while restyling of existing or newly generated images leverages techniques such as img2img guided by ControlNet with Stable Diffusion. These nodes implement crucial feedback loops \textbf{(D2.5}) through direct manipulation and explicit actions like ``Approve'', which signals the backend to persist the current state of the node's content.

\textbf{Workflow Execution and Caching}: To ensure responsiveness, especially in complex workflows with multiple interconnected infomorphs, the backend implements a caching and selective recomputation strategy. The workflow defined on the canvas is represented internally as a Directed Acyclic Graph (DAG). We utilize Python's NetworkX library to manage this graph structure, and translate between xyflow's representation, and our internal representation, and track dependencies between nodes. When a user modifies a node's configuration, input data, or provides feedback (like editing content in a Viewer), the backend identifies only the subset of downstream nodes affected by this change. It then triggers recomputation for only those nodes, reusing cached results for all unaffected upstream nodes. This significantly reduces latency and computational cost compared to re-running the entire workflow on every change.
\section{Discussion and Future work}

While the scenario walkthrough illustrates the potential benefits of this approach—particularly in enhancing transparency and control compared to \textit{monolithic} and \textit{opaque} systems of today, several important points warrant discussion and highlight avenues for future work.
\vspace{-4pt}
\subsection*{Scope and Applicability} 

Our proposed design space and the \systemname prototype deliberately center on information work. These tasks primarily involve the gathering, synthesis, restructuring, and transformation of content from existing sources into new document forms like reports, presentations, or structured summaries. We acknowledge that this scope is a subset of the broader landscape of knowledge work. Consequently, the proposed concepts and system may be less directly applicable to tasks where the core activity is, for example, creative writing and brainstorming, intricate mathematical modeling, scientific data analysis, or complex software development, although elements of information synthesis exist in these. Even within information work, evaluating the practical boundaries and refining the applicability of such an interaction style will necessitate longer-term user studies involving diverse knowledge workers and tasks to systematically identify where \textit{infomorph}-driven workflows excel and where they encounter limitations or fail.
\vspace{-4pt}
\subsection*{Canvas Metaphor}

The scenario in the paper demonstrates how \systemnamedot's canvas interface can provide transparency into the transformation process and allow users to construct complex workflows by chaining infomorphs. Compared to single-shot generation or purely chat-based iteration, this visual approach offers more user control and awareness of the intermediate steps. While the canvas proved useful for demonstrating the concepts of the design space, it is not without its limitations. Canvas interfaces inherently demand significant screen real estate and can quickly become visually cluttered and challenging to navigate as the number of nodes and connections increases. A scalability issue is hinted at even by the large figures required to depict the modest workflows within this paper. Maintaining focus-context tradeoffs can become cumbersome. Therefore, determining optimal interaction techniques for flexible, multimodal, AI-augmented information transformation workflows continues to be a key area for future work. Future research should explore and evaluate alternative interaction paradigms, structured list-based or outline-based workflow editors, tighter integration of generative AI features within the context of source or target documents, or perhaps adaptive interfaces that auto simplify based on user expertise or task complexity.

\subsection*{App-less future or Better information glue?} 
We emphasize that neither the design space proposed here nor the \systemname prototype aims to be a standalone, ``one app for everything'' solution. While it's true that the rapid advancements and broad multimodal capabilities of Generative AI models are beginning to challenge the traditional ecosystem of specialized applications (like the MS Office suite or Adobe Photoshop). A single powerful model can now perform an impressive range of tasks such as summarization, translation, content generation, image creation, and basic analysis. Previously, this might have required multiple distinct tools and software. However, there is a long way to go between current AI capabilities and the nuanced reliability, controllability, and domain-specific understanding required for many complex, high-stakes knowledge work tasks. This is especially true in scenarios that benefit and thrive from human-in-the-loop interactions. Furthermore, specialized applications often persist because they offer crucial efficiency advantages through highly optimized user interfaces, deeply integrated features tailored to specific user workflows, or performance benefits for core tasks. 

This is seen with \systemname which is powered by various Generative AI techniques for tasks like content extraction, summarization, text generation, and image generation and editing. However, it explicitly operates on inputs from and generates outputs compatible with these established, specialized application formats such as Microsoft Excel (.xlsx), PowerPoint (.pptx), PDF (.pdf), and Word documents (.docx). This design acknowledges and embraces the continued utility of these conventional applications that many people use, and continue to use for everyday work. At least for the near future, this evolving landscape, likely featuring a hybrid ecosystem composed of both powerful Generative AI capabilities and enduring specialized applications. 

The difficulty of moving and transforming information between diverse tools remains a critical bottleneck in knowledge work, driven by incompatible formats, structures, modalities (text, image, table), and high cognitive load for manual conversion. This friction is evident even with advanced AI like Office CoPilot or ChatGPT. Our work tackles this ``information glue'' problem by proposing modular \textit{infomorphs} and a guiding design space to think and reason about user control and agency. These conceptual tools are intended to foster the creation and systematic evaluation of novel interfaces and interaction techniques. The ultimate goal is to find approaches, perhaps distinct from \systemnamedot, that achieve a more seamless and controllable alignment between Generative AI output and specific user intentions during complex synthesis tasks.
\vspace{-4pt}
\section{Conclusion}

In this paper, we introduced a framework for interactive, multimodal document synthesis, featuring a five-dimensional design space and the core concept of \textit{infomorphs}, which are modular, user-steerable, AI-powered information transformations. These provide a structured approach for leveraging Generative AI effectively in document-centered knowledge work. We presented \systemnamedot, a canvas-based prototype, demonstrating how composing \textit{infomorphs} can enable flexible, user-steerable workflows across diverse multimodal information sources. Ultimately, this work makes an argument for, and offers thought tools to build more transparent, user-steerable AI systems for document-centric knowledge work.

\begin{acks}
We thank Andy Wilson (Microsoft Research) for his valuable feedback across multiple iterations of DocuCraft. We also thank our colleagues-Anjan Goswami (Microsoft PowerPoint), Sumithra Bhakthavatsalam, Gongjie Qi, and Gaurav Tendolkar (Microsoft Designer)-whose deep knowledge of Microsoft file systems was valuable in the development of DocuCraft.
\end{acks}

\bibliographystyle{ACM-Reference-Format}
\bibliography{references}

\end{document}